\documentclass[aps,prb,twocolumn,showpacs,showkeys]{revtex4}
\usepackage{graphicx}
\usepackage{amsmath}
\usepackage{hyperref}

\begin{document}

\title{The Sznajd dynamics on a directed clustered network\footnote{dedicated to Dietrich Stauffer for his 65-th birthday}}

\author{K. Malarz}
\homepage{http://home.agh.edu.pl/malarz/}
\email{malarz@agh.edu.pl}
\affiliation{
Faculty of Physics and Applied Computer Science,
AGH University of Science and Technology,\\
al. Mickiewicza 30, PL-30059 Krak\'ow, Poland.
}

\author{K. Ku{\l}akowski}
\email{kulakowski@novell.ftj.agh.edu.pl}
\affiliation{
Faculty of Physics and Applied Computer Science,
AGH University of Science and Technology,\\
al. Mickiewicza 30, PL-30059 Krak\'ow, Poland.
}

\date{\today}

\begin{abstract}
The Sznajd model is investigated in the directed Erd{\H o}s--R\'enyi network with the clusterization coefficient enhanced to 0.3 by the method of Holme and Kim (Phys. Rev. {\bf E65} (2002) 026107).
Within additional triangles, all six links are present.
In this network, some nodes preserve the minority opinion.
The time $\tau$ of getting equilibrium is found to follow the log-normal distribution and it increases linearly with the system size.
Its dependence on the initial opinion distribution is different from the analytical results for fully connected networks.
\end{abstract}



\maketitle

%% ##########################################################################
\section{Introduction}
%% ##########################################################################

As it was formulated by Woodrow Wilson, opinion ultimately governs the world \cite{wood}; the common interest in its research is then more than justified. 
The rule of the public opinion (PO) got a new dimension when the mass media appeared at the beginning of the 20-th century. The change was recognized by Walter Lippmann who shaped our understanding of PO \cite{lip}. In social sciences, large effort is devoted to capture necessary ingredients of a qualitative description of PO \cite{shsh}. However, the role of structure of the social network was also thoroughly investigated \cite{wasfa}. The main message from the social sciences is that the PO dynamics proceeds in a landscape
of established identifications. Then, variations of PO are possible through local current reinterpretations of political events rather 
than through converting of large groups \cite{ww}. Important difference --- the so called pluralistic ignorance --- appears between the actual 
and the expressed opinion, and it is only the latter which influences the dynamics \cite{nn}. 

Physicists entered to the field via statistical mechanics \cite{dss}; a recent review on this kind of research of PO dynamics can be found in Ref. \onlinecite{caflo}. In these quantitative models, main role is played by the structure of social networks. Actual shape of this structure is a matter of discussion \cite{schnegg} and simple solutions cannot be expected. However, various models have been and continuously are constructed to capture this or that feature of the structure of at least selected social groups. The voter model describes the opinion spreading via links of direct neighborhood \cite{vot,red}. In the Sznajd model \cite{szn,sz2,sz3} an accordance of two neighbors is necessary to influence other neighbors of the pair. 
In the Deffuant model \cite{weis}, attention is paid to the difference of opinion of the neighbors; if this is too large, the transfer of opinion is blocked. These efforts seem to complete a consistent path towards a realistic although model description of the PO dynamics within the social structure approximated by a network. 
 
The aim of this work is to make next step in this direction. The choice we have to make is twofold: to select the dynamic rule and to select the structure. Here we use the Sznajd model; in our opinion this model gives a bridge from the Granovetter's concept of threshold \cite{gra} to computer simulations. Here the threshold is set as two persons, what is an important generalization of the voter model and what allows to take into account the correlations of opinions of neighbors. The structure of our model network is random and clustered \cite{wdn}, it shows the small-world property \cite{wdn} and it contains directed links; the last condition comes from the observed asymmetry of social relations \cite{gon}. Below in the text, the construction of the network is described in detail.

Here we concentrate on the calculations of the time $\tau$ of getting a stationary distribution of opinions. In this final stage although local opinions can change, the average opinion calculated over the whole network remains constant. In political sciences, the initial support for a new government collapses in some time, which can be identified by $\tau$. To give examples from recent Polish history, the timescale of this fall seems to be about two years (Jerzy Buzek) or shorter (Leszek Miller) \cite{cbos1}. We note that our calculations do not capture the metastable states of PO, where the majority is silent and activates only during an open debate \cite{nn}.

%% ##########################################################################
\section{The network and the model}
%% ##########################################################################

\begin{figure}
(a) \includegraphics[scale=0.25]{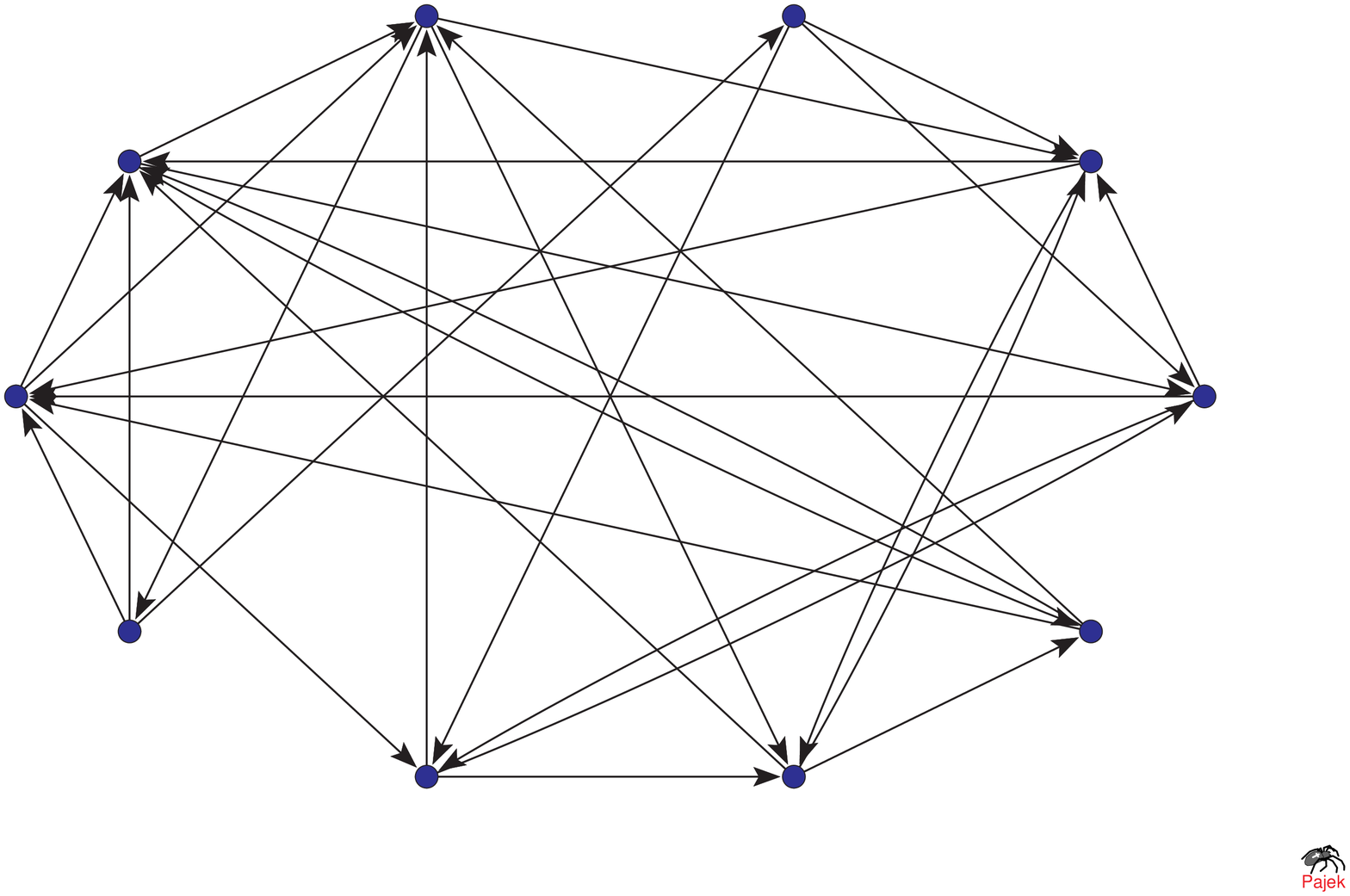}\\
(b) \includegraphics[scale=0.25]{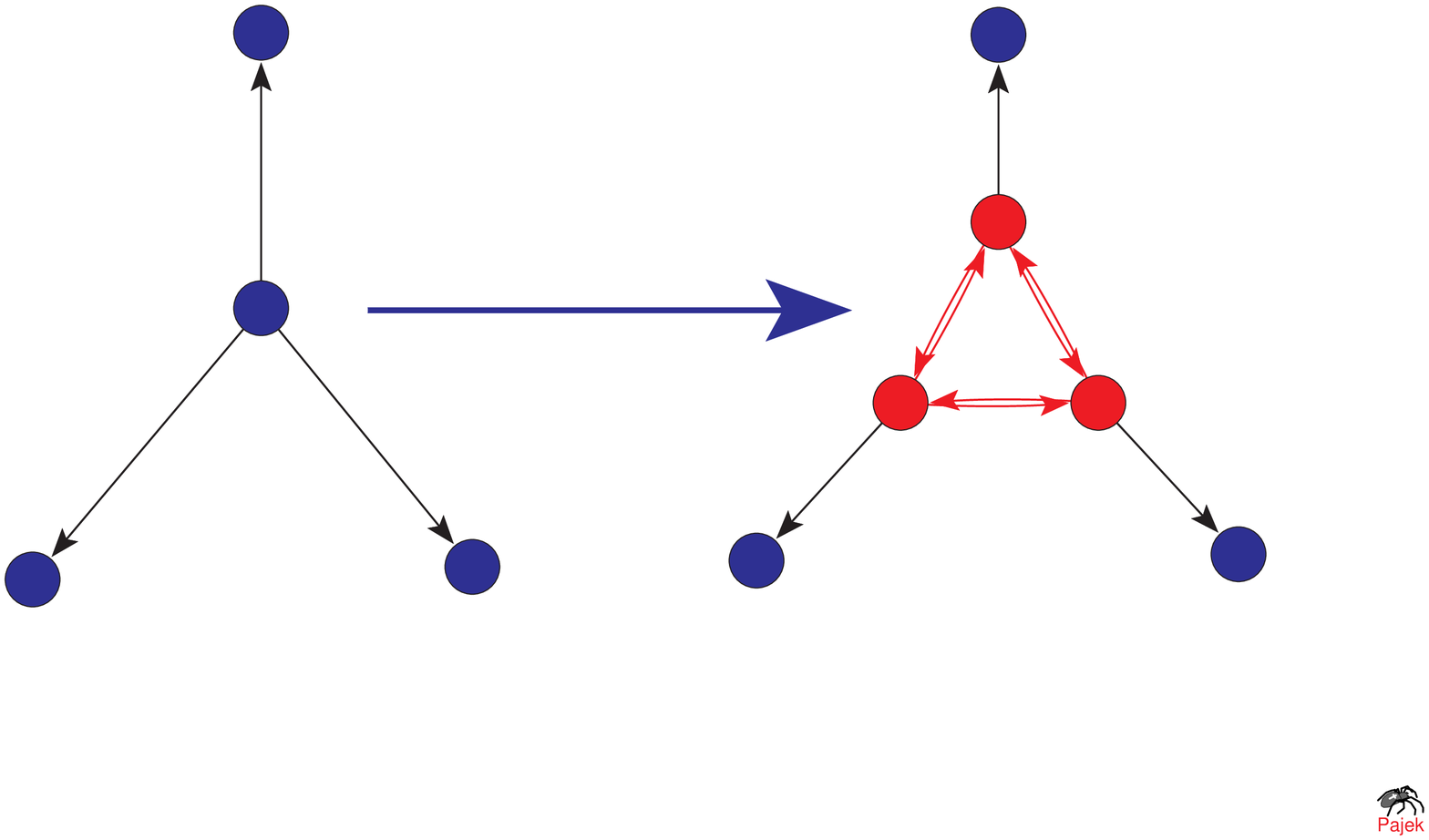}\\
(c) \includegraphics[scale=0.25]{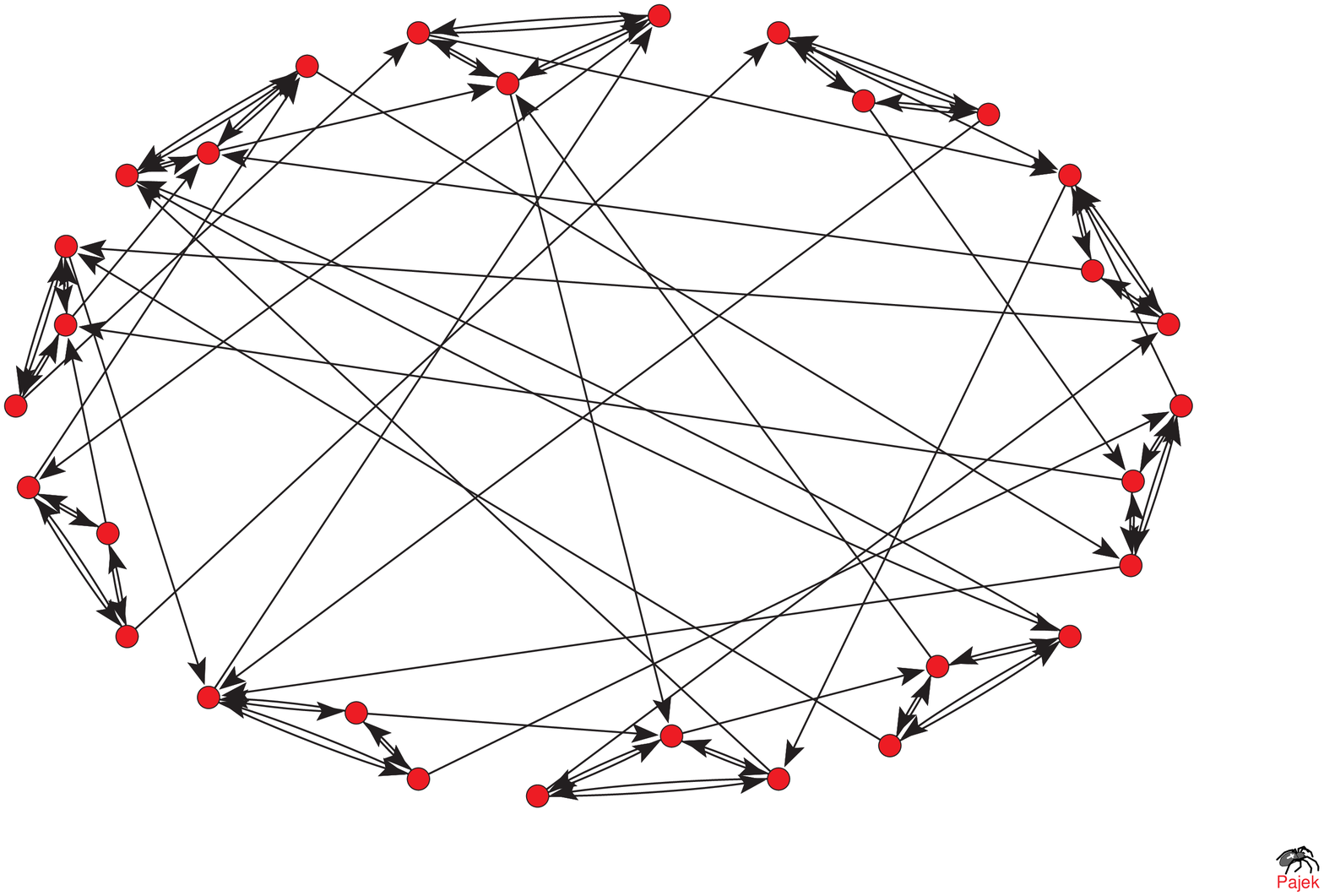}
\caption{\label{fig-net} Network construction.
 In both cases $P_{\text{out}}(k)=\delta_{k,3}$. 
$N=30$.}
\end{figure}

The model society is the directed network \cite{gon} based on the Erd{\H o}s--R\'enyi random graph \cite{CRG} with enhanced clusterization \cite{holm}.
The primary network is constructed such that each node has exactly three out-going links, i.e.
$P_{\text{out}}(k)=\delta_{k,3}$,
as presented in Fig. \ref{fig-net}(a).
The second step in network preparation is a splitting of each node of the primary network into the well clustered set of three nodes presented in Fig. \ref{fig-net}(b). 
Other method of control the network topology was presented in Ref. \onlinecite{serrano}.
An example of the final network of $N=30$ nodes is presented in Fig. \ref{fig-net}(c).

%% --------------------------------------------------------------------------
%% --------------------------------------------------------------------------

The spin like opinion $s_i=\pm1$ is assigned to each node/actor $i$.
The opinion dynamics is governed by original Sznajd rule \cite{szn}: each time step we scan all $N$ nodes according to a random permutation of the nodes labels.
For each investigated node $i$ one of its three out-neighbors $j$ is selected.
If the opinions of such selected pair are identical ($s_i s_j=1$) then
each out-neighbors of the pair accepts the pair's opinion with the probability $p$. 

Initially, the fraction $p_0$ is of the negative opinion.
The results of simulations are averaged over $N_{\text{run}}$ independent runs.

%% ##########################################################################
\section{Results}
%% ##########################################################################

For the Sznajd model, the system tends to the state of uniform opinion, i.e.
\[ m=\sum_{i=1}^N s_i=\pm 1, \]
basing on the majority of the initial opinions.
Here however, some actors will never change their initial opinion and full consensus may be never reached. 
Let us consider a triangle of actors presented in the right part of Fig. \ref{fig-net}(b).
In this case, when all red spins have the same initial opinion, they will sustain it and will support each other in keeping it constant.
The fraction of such triangles depends on the concentration of initial majority $p^c_0=\frac{1}{2}-|p_0-\frac{1}{2}|$.
The simulation takes $N_{\text{iter}}$ steps (and in each time step $N$ actors pairs is investigated) unless the saturation opinion 
\[
m_\infty=
\begin{cases}
2e^{-3}(1+2p^c_0)(1-p^c_0)^2-1 \iff p_0>0.5\\
1-2e^{-3}(1+2p^c_0)(1-p^c_0)^2 \iff p_0<0.5
\end{cases},
\]
$(|m_\infty|<1)$ is reached earlier.

\begin{figure}
(a) \includegraphics[scale=0.6]{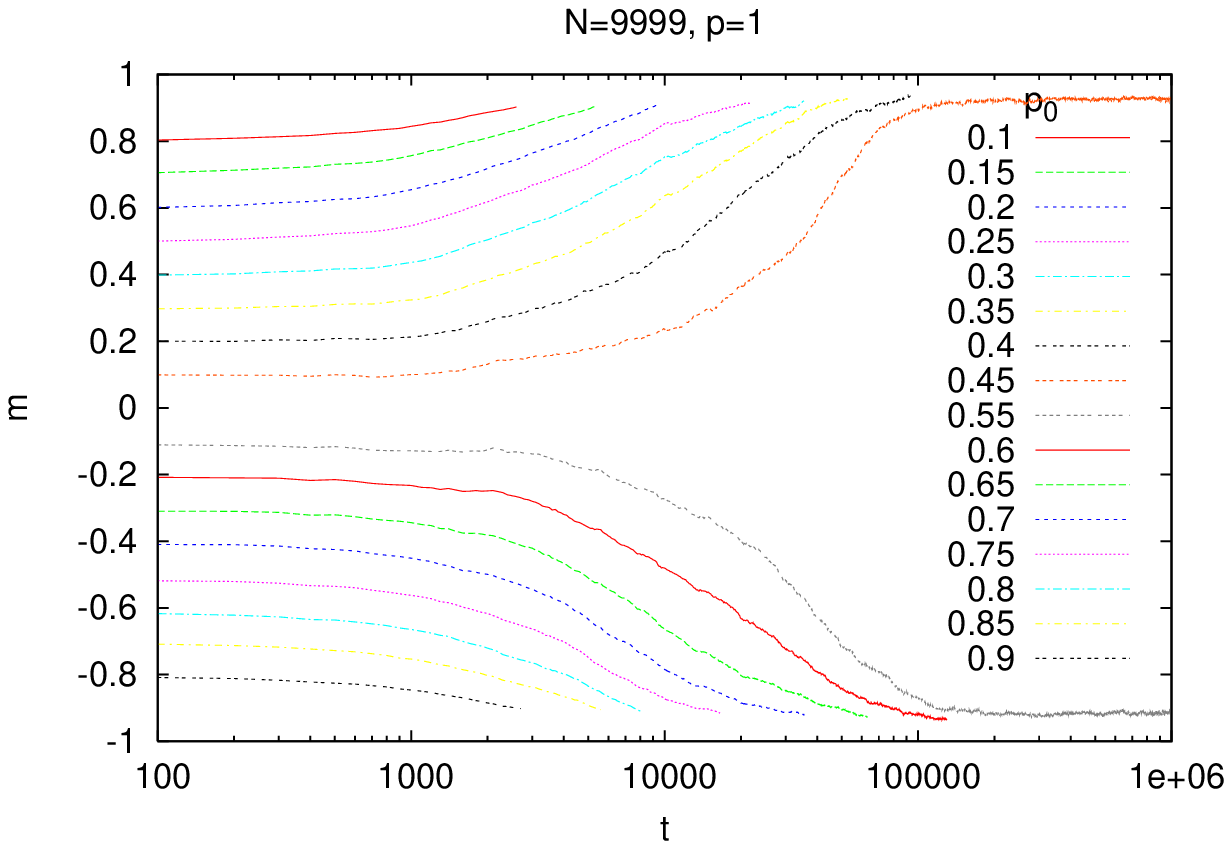}\\
(b) \includegraphics[scale=0.6]{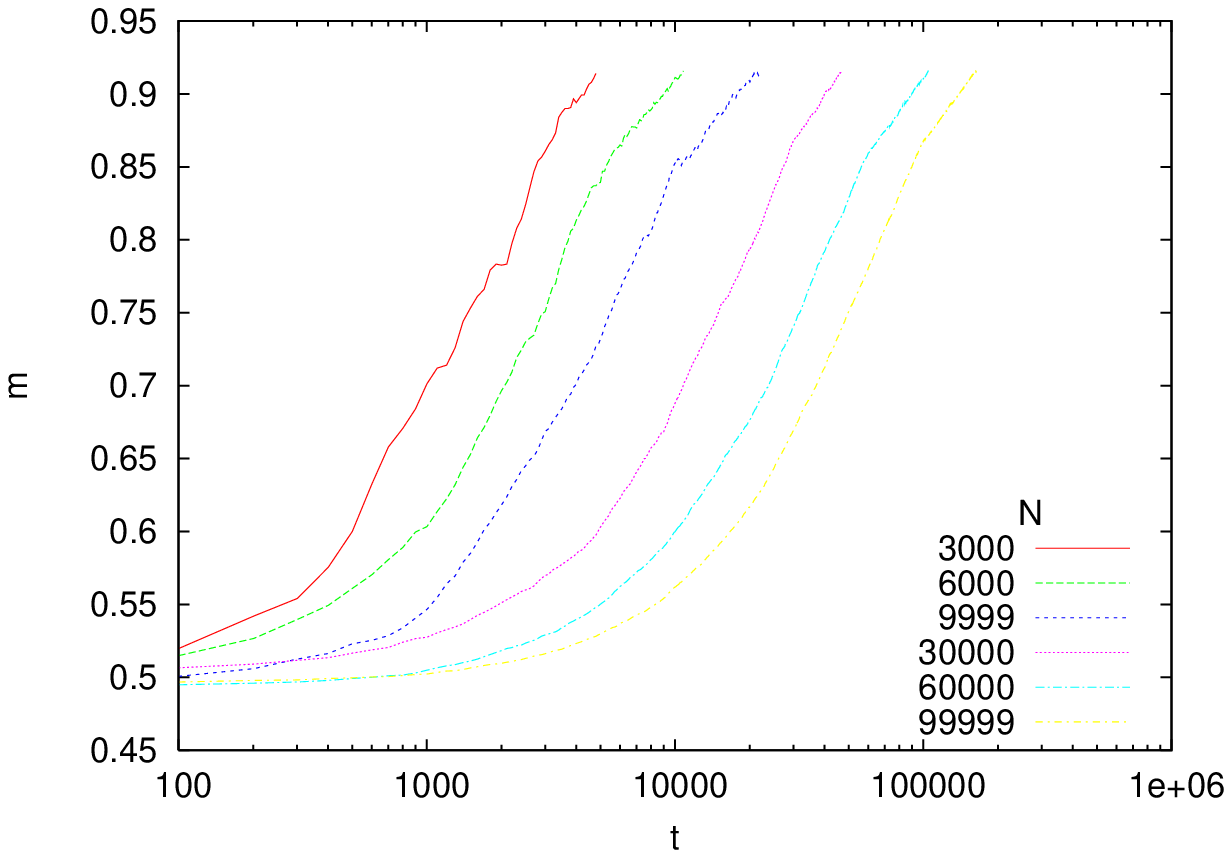}\\
(c) \includegraphics[scale=0.6]{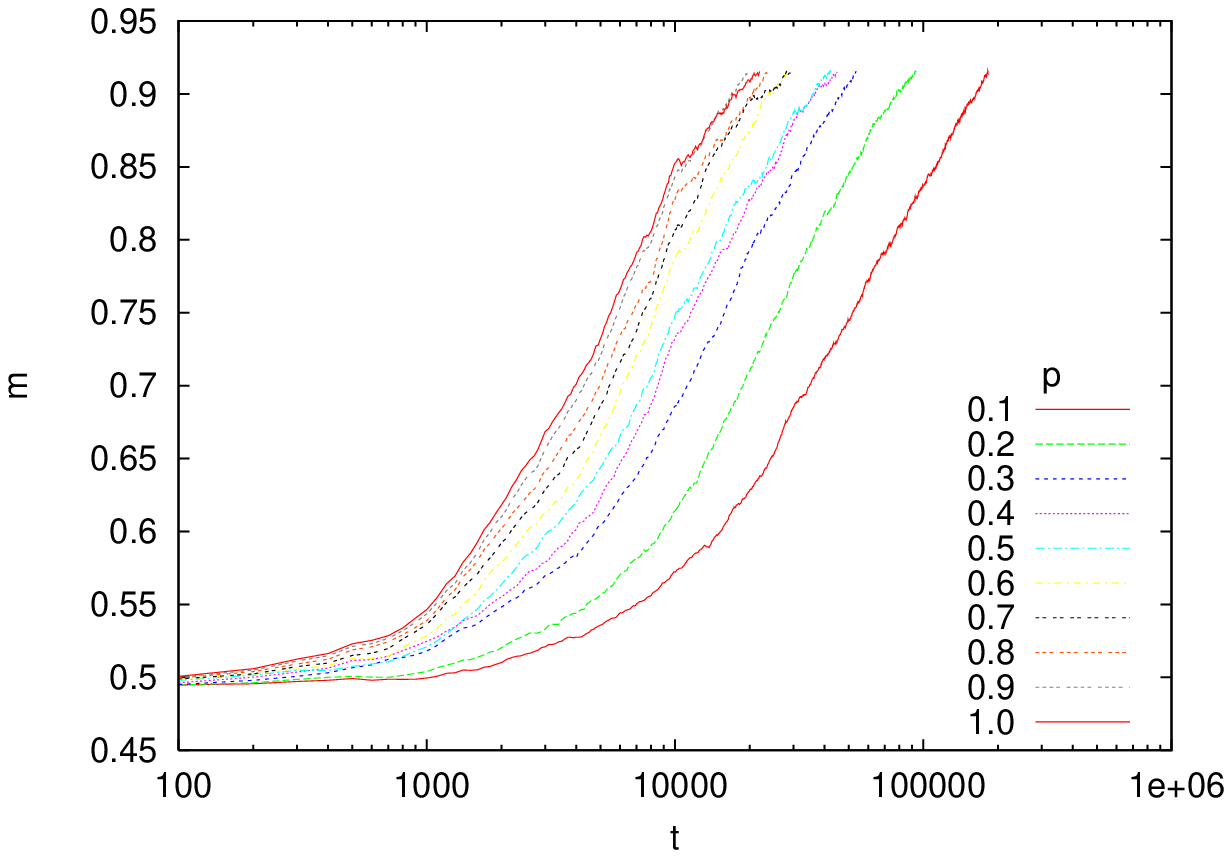}
\caption{\label{fig-mvst} Time evolution of average opinion dynamics for different values of
(a) initial fraction $p_0$ of one of the opinion,
(b) system size $N$
and (c) probability of following Sznajd rule $p$.}
\end{figure}

The time evolution of the average opinion $m$ for 
\begin{itemize}
\item different values of initial fraction $p_0$ of one opinion $p_0$,
\item system size $N$
\item and probability of following Sznajd rule $p$
\end{itemize}
are presented in Fig. \ref{fig-mvst}(a), (b) and (c), respectively.

%% --------------------------------------------------------------------------
%% --------------------------------------------------------------------------

\begin{figure}
(a) \includegraphics[scale=0.6]{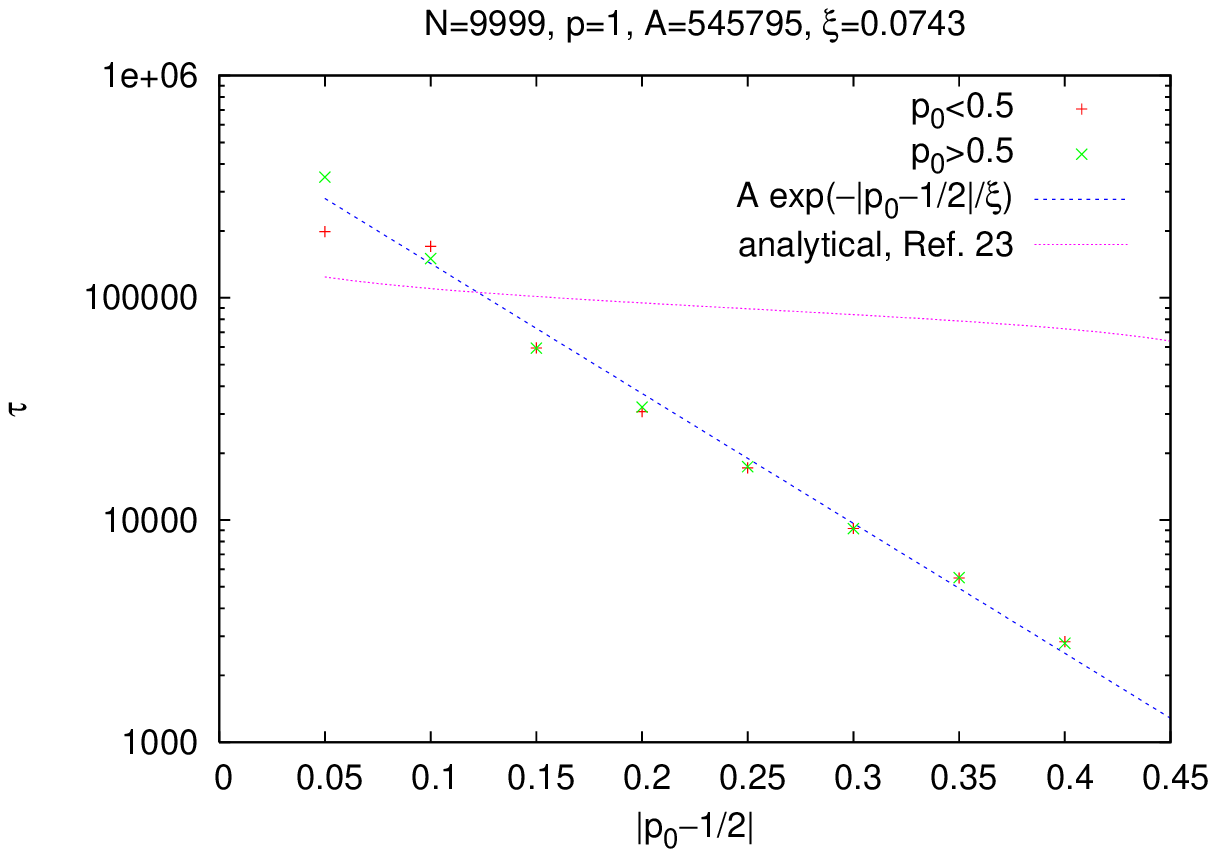}\\
(b) \includegraphics[scale=0.6]{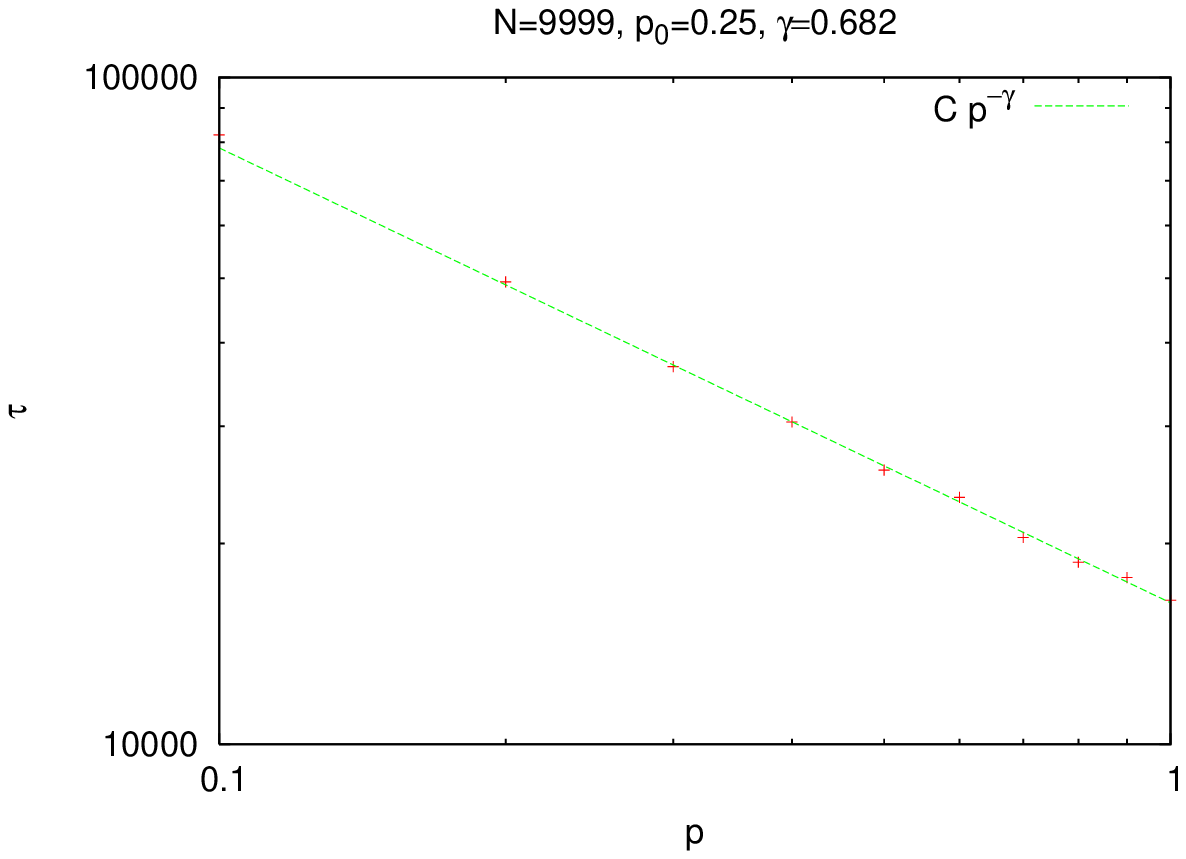}\\
(c) \includegraphics[scale=0.6]{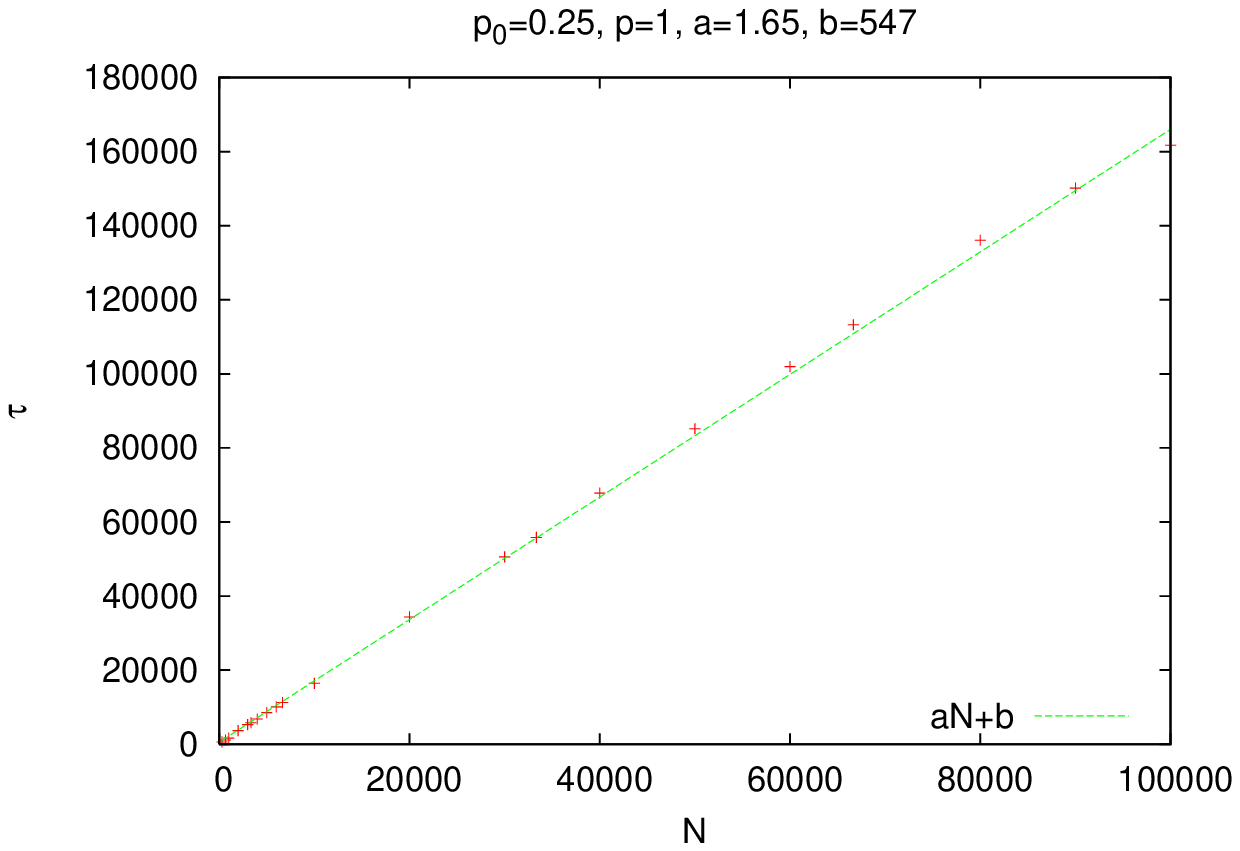}
\caption{\label{fig-tau} Time $\tau$ of reaching the consensus for different values of 
(a) initial fraction $p_0$ of one of the opinion,
(b) system size $N$
and (c) probability of following Sznajd rule $p$, when the other parameters are fixed on $N=9999$, $p=1$, $p_0=0.25$.
The fitted parameters are $A=545795$, $\xi=0.0743$, $a=1.65$, $b=547$, $C=16294$, $\gamma=0.682$.}
\end{figure}

The average time $\tau$ of reaching saturation opinion $m_\infty$ depends
\begin{itemize}
\item exponentially with difference $p_0$ from the critical concentration 1/2: 
\[\tau \propto \exp\dfrac{-|p_0-1/2|}{\xi},\]
with $\xi=0.0743$
(Fig. \ref{fig-tau}(a)),
\item according to power law with the probability of following Sznajd rule
$\tau\propto p^{-\gamma}$, $\gamma=0.682$
(Fig. \ref{fig-tau}(b))
\item and linearly with the system size 
$\tau = 1.65N + 547$
(Fig. \ref{fig-tau}(c)).
\end{itemize}
The last result was observed and explained earlier in Ref. \onlinecite{slal} as well.

%% --------------------------------------------------------------------------
%% --------------------------------------------------------------------------

\begin{figure}
(a) \includegraphics[scale=0.6]{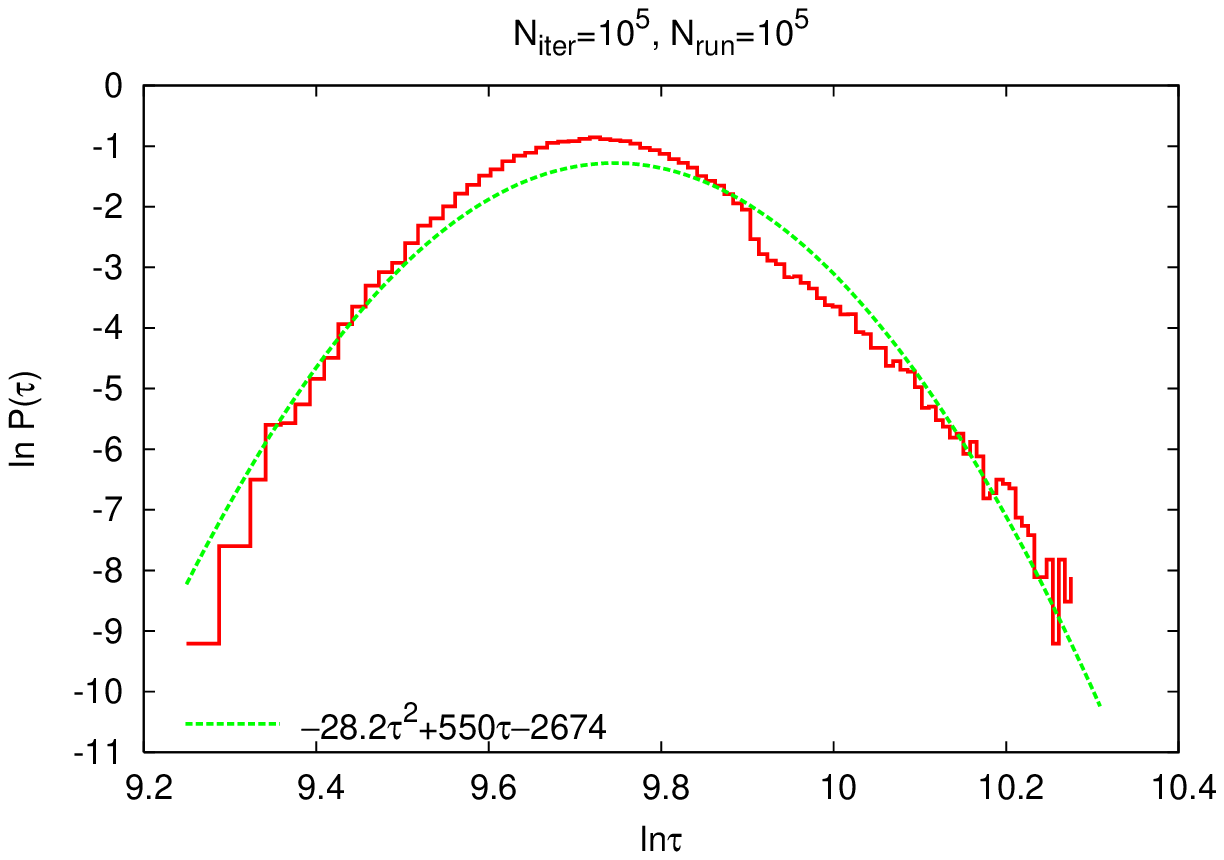}\\
(b) \includegraphics[scale=0.6]{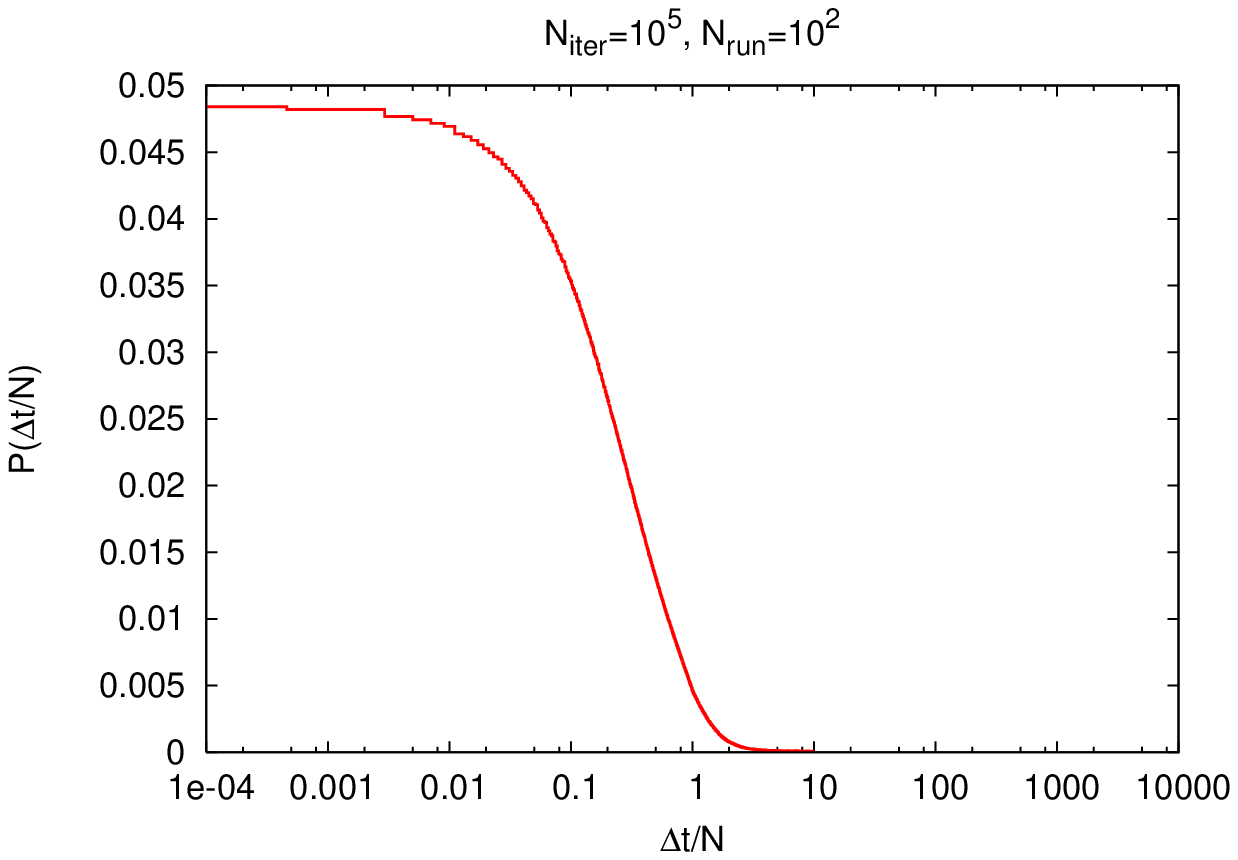}\\
(c) \includegraphics[scale=0.6]{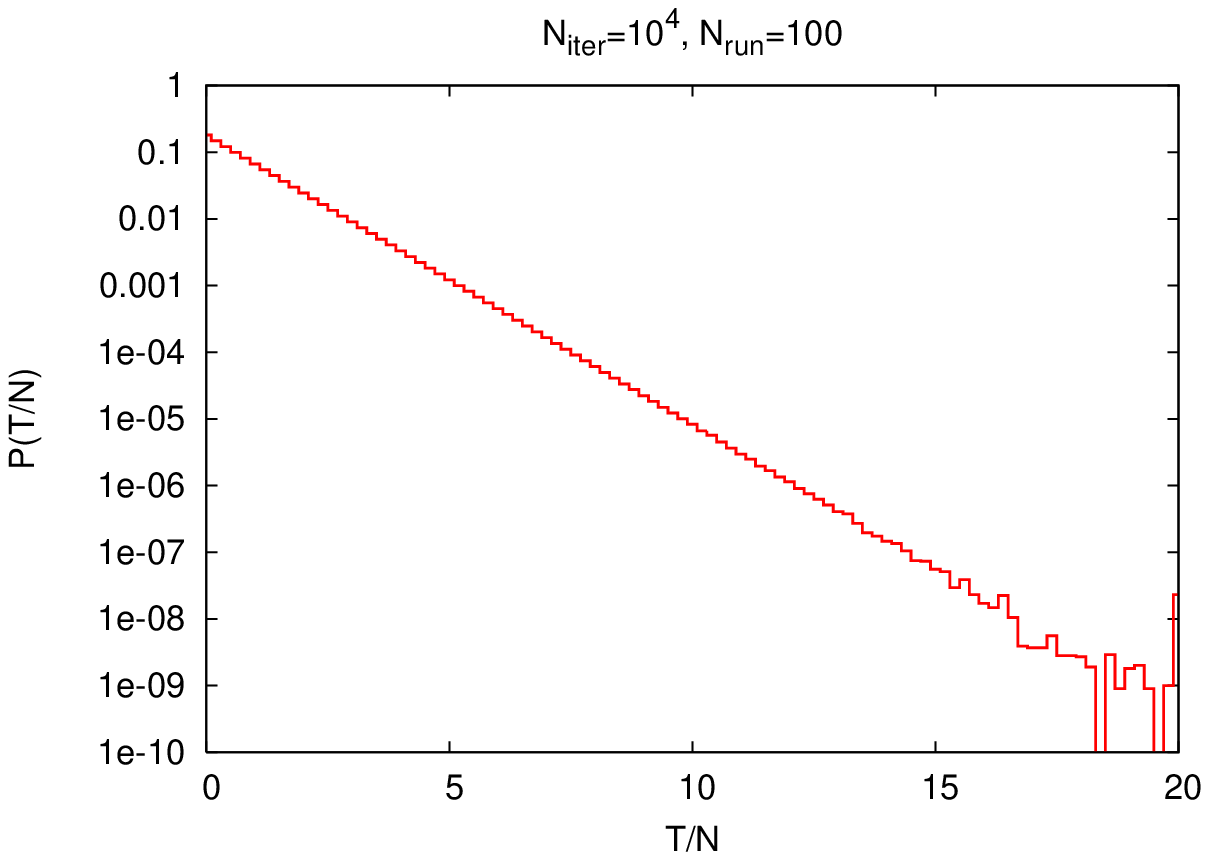}
\caption{\label{fig-tau-dt-T} The distribution of time
(a) $\tau$ of reaching the consensus,
(b) $\Delta t$ of delay between subsequent flips of the given spin,
(c) $T$ of visiting the given spin.
In the latter case the spins are visited in completely random fashion and not according the permutation of the spin label.
$N=9999$, $p_0=0.25$, $p=1$.}
\end{figure}

The distribution $P(\tau)$ of reaching the consensus time seems to be log-normal.
An example of such distribution for $N_{\text{run}}=10^5$ independent simulations is shown in Fig. \ref{fig-tau-dt-T}(a).
We have checked the distribution $P$ of times $\Delta t$ between subsequent flips of given spins and the one of times $T$ between reaching the same node.
The distributions are presented in Figs. \ref{fig-tau-dt-T}(b) and  \ref{fig-tau-dt-T}(c). 
In the latter case nodes are selected in purely random fashion and not according to actors labels permutations.

%% --------------------------------------------------------------------------
%% --------------------------------------------------------------------------

%% ##########################################################################
\section{Discussion}
%% ##########################################################################

As remarked above, in our network some nodes have no in-links.
 These nodes and some of their neighbors remain in the initial state, and therefore full accordance
of opinions cannot be reached. This is an important difference between our model and other simulations of the Sznajd dynamics.

%% --------------------------------------------------------------------------
\begin{figure}
\includegraphics[scale=0.6]{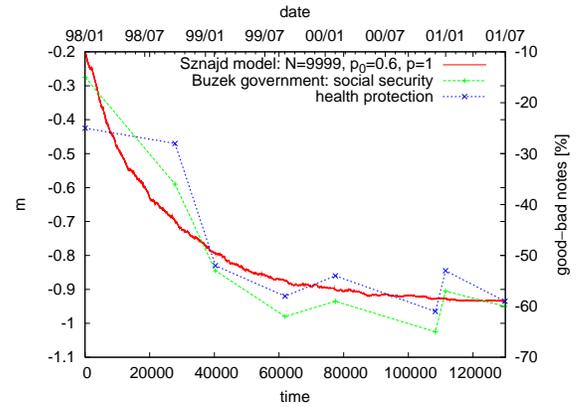}
\caption{\label{fig-sznajd-buzek} The fall of the public support for the policy of the government of Jerzy Buzek on the social issues and the time evolution of the main opinion in Sznajd model for $N=9999$, $p_0=0.6$ and $p=1$.
}
\end{figure}
%% --------------------------------------------------------------------------

The shape of the obtained curve on the mean opinion against time agrees qualitatively with the accessible experimental data \cite{cbos1} on the fall of the public support for the policy of the government of Jerzy Buzek on the social issues: social security and health protection between January 1998 and June 2001. This comparison is shown in Fig. \ref{fig-sznajd-buzek}. The curves on other issues are practically the same (agricultural policy), show the similar timescale (economy and management) or do not vary significantly (foreign affairs). 

The linear dependence of the time $\tau $ on the system size $N$ is the same as found analytically for the fully connected graph \cite{slal}.
On the other hand, the theoretical prediction \cite{slal} on the $\tau$ dependence on $p_0$ is not confirmed by our simulations. This means, that the result $\tau \propto N$ is valid for a larger class of networks.

In our simulations, all nodes are selected at each time step, in different order, and the Sznajd rule is executed with probability $p$. The $\tau$ dependence on $p$ is slower than just $1/p$; the latter could be expected if the decrease of $p$ is equivalent just to slowing the process down. The difference can be due to the fact, that once $p<1$, some nodes are not updated at a given time step. Paradoxically, the presence of these temporal ``contrarians'' seems to accelerate the process of getting stationary state. 

The obtained distribution of $\tau$ is close to the log-normal distribution, in accordance with the discussion in Ref. \onlinecite{ssmo}. This means that the phase transition is well defined. The distribution of times between subsequent changes of state of a given node is shown in Fig. \ref{fig-tau-dt-T}(b). The plot is different from the scaling $\tau ^{-3/2}$, obtained in Ref. \onlinecite{szn} for the one-dimensional system. 

Once $p=1$, the distribution of time between subsequent picking up of the same node (irrespectively whether it is changed or not) is the tent function, with the average $\tau = N$. In the case when the nodes are selected randomly and not as noted above, the same distribution is the exponential one, as in Fig. \ref{fig-tau-dt-T}(c). 

Concluding, the Sznajd dynamics has been applied to the directed network designed as to reproduce the small world effect and clusterization, as in social networks. The results on the time $\tau$ of getting stationary state are compared to the analytical calculations for the fully connected network \cite{slal}. 
Although the dependence of $\tau$ on the initial state is different than in theory, the system size dependence of $\tau$ is the same.

The message for the social applications is that the time of attaining the stationary distribution increases linearly with the system size. Among other consequences, this means that an initial support for a new government persists twice longer in a twice larger country; larger countries have more time for reforms. Still, even if the initial state is the longest-living $p_0=0.5$, this time is not infinite, as it is shown in Fig. \ref{fig-tau}(a). As remarked in the Introduction, all that is valid as long as opinions are adjusted in small groups. Important political events can reveal a metastable character of the opinion distribution; however, this possibility remains out of frames of this work.

%% ##########################################################################
\begin{acknowledgments}
The numerical calculations were carried out in ACK\---CY\-F\-RO\-NET\---AGH.
The machine time on HP Integrity Superdome is financed by the Polish Ministry of Science and Higher Education under grant No. MNiI/\-HP\_I\_SD\-AGH/\-047/\-2004.
\end{acknowledgments}
%% ##########################################################################

%% ##########################################################################

\end{document}